\input harvmac

%%%% definitions %%%%%%

\def\p{{\partial}}

\def\g {{\gamma}}

\def\e {{\epsilon}}

\def\l{\lambda}
\def\th{\theta}
\def\g{\gamma}
\def\G{\Gamma}
\def\e{\epsilon}

\def\be{\begin{eqnarray}}
\def\ee{\end{eqnarray}}

\centerline{{\bf Non-critical strings and ${\cal W}$-algebras}  }
\bigskip \centerline{{\bf  Ram
Sriharsha}\footnote{$^{\dag}$}{ harsha@glue.umd.edu}}
\centerline{\it Department of Physics, University of Maryland}
\centerline{\it College Park, MD 20742-4111}

\bigskip
\bigskip
\baselineskip 18pt
\bigskip
\noindent

Singularities of Spin(7) manifolds are considered in the worldsheet
approach, and it is argued that the internal CFT describing a
singular spin(7) has an enhanced ${\cal SW}({3\over 2},{3\over
2},2)$ algebra tensored with ${\cal N}=1$ linear dilaton CFT, much
like singular Calabi-Yau CFTs have a ${\cal N}=2$ linear dilaton
tensored with a suitable ${\cal N}=2$ SCFT. Upon adding fundamental
strings, these vacua are related to $AdS_3$ vacua with ${\cal N}=1$
supersymmetry, completing the worldsheet classification of $AdS_3$
vacua with NS flux.

 \Date{14th November,2006}

\newsec{Introduction}

It is well known that the starting point for the description of
Calabi-Yau compactifications in the RNS approach involves ${\cal
N}=2$ SCFTs with a charge integrality condition that allows GSO
projection to preserve space-time supersymmetry. The other
compactifications that can be similarly treated in RNS formalism are
the spin(7) and $G_2$ holonomy spaces. In this paper, we consider
spin(7) holonomy manifolds. It has been shown \ref\sv
{S.~L.~Shatashvili and C.~Vafa, ``Superstrings and manifold of
exceptional holonomy,'' arXiv:hep-th/9407025} that the data
describing a spin(7) compactification is the ${\cal SW}({3\over
2},2)$ algebra with $c =12$. This allows us to define GSO projection
and preserve two supercharges in the uncompactified directions of
${\bf R}^2$. We will review the derivation of this enhanced
superconformal algebra in section 2, following an approach that was
used in the case of Calabi-Yau compactifications. In the case of
Calabi-Yau spaces it is well known that the CFT describing the
compactification becomes singular as certain K{\"a}hler and/ or
complex structure parameters are varied, and the singular CFT that
one typically gets is based on the ${\cal N}=2$ linear dilaton CFT
tensored with a suitable internal SCFT with ${\cal N}=(2,2)$
supersymmetry (the details of this CFT depend on the nature of the
singularity). A natural generalization is to try and determine the
type of singularities that a spin(7) manifold can have, and describe
the physics near this singularity in terms of a Liouville type
theory. The straightforward approach to this question seems
complicated because not much seems to be known about the moduli
spaces of spin(7) holonomy manifolds, and even less is known about
the type of singularities that occur at a finite distance in moduli
space. One can however note that the singular CFT is well described
by a non-critical string, and that non-critical strings can be
viewed as critical strings if we include the Liouville mode, so we
are in essence looking for non-critical strings that contain a
Liouville type mode, and further contain a ${\cal SW}({3\over 2},2)$
SCA. As we show section 3, this leads naturally to the class of
non-critical strings containing a linear dilaton CFT and a ${\cal
{SW}}({3\over 2},{3\over 2},2)$ SCA. Alternatively, one can start
with these non-critical backgrounds, and add fundamental strings
that span ${\bf R}^2$. Upon including back-reaction, one expects to
end up with an $AdS_3 \times {\cal N}$ background, where ${\cal N}$
is a suitable ${\cal N}=(1,1)$ SCFT that allows string theory to
preserve two supercharges in $AdS_3$. For weakly coupled sigma
models, we will show using supergravity that these backgrounds can
preserve ${\cal N}=1$ space-time SUSY only if there exists on the
worldsheet a spin-${3\over 2}$ operator that is holomorphic. The
minimal ${\cal N}=1$ SCA that includes this spin-${3\over 2}$
operator is nothing but the ${\cal SW}({3\over 2},{3\over 2},2)$
algebra. ${\cal N}=1$ ${\cal SW}({3\over 2},{3\over 2},2)$ algebras
have been recently discussed in a paper by Noyvert \ref\Noyvert{
  B.~Noyvert,
  ``Unitary minimal models of SW(3/2,3/2,2) superconformal algebra and
  manifolds of G(2) holonomy,''
  JHEP {\bf 0203}, 030 (2002)
  [arXiv:hep-th/0201198].}, to which we refer the reader for details.

\newsec{Compact Spin(7) manifolds}
We will consider Type II superstrings propagating in a geometry of
the form ${\bf R^{1,1} \times X_8}$, where the internal manifold is
taken to correspond to an ${\cal N}$ = 1 SCFT on the world-sheet,
and will work in the RNS formalism. Then, it is well known how to
construct the space-time supersymmetry vertex. In other words, one
would like to know how to write down the target space-time
supercharge, in terms of the world-sheet fields, so that space-time
supersymmetry can be made explicit in the RNS version. The vertex
operator for the space-time supercurrent is, in the (-1/2) picture
given as:

\eqn\ai{ {\cal V}_{\alpha} = {{e}^{-{\phi}/2} } S_{\alpha} {\Sigma}
}

where ${\phi}$ is the bosonized super-ghost, S$_{\alpha}$ is the
spacetime spin field, while the field ${\Sigma}$ is the spin field
coming from the Ramond sector of the internal SCFT to which the
compactification on ${\bf X_8}$ corresponds.

An explicit formula for the spacetime spin field S, is given by
bosonization \ref\p{J.~Polchinski, ``String theory. Vol. 2:
Superstring theory and beyond,''}. Using the known OPEs of the
fields ${\phi}$ and S, and insisting on spacetime supersymmetry, one
easily arrives at the following OPEs for the fields ${\Sigma}$.

\eqn\aii{{\Sigma(z)} {\Sigma(0)}  {\sim}     {1\over z} + O(z) }

In order for the space-time supersymmetry vertex ${\cal V}_{\alpha}$
to give rise to a supercharge, the vertex must be an operator of
dimension 1, and this forces the dimension of ${\Sigma}$ to be 1/2.
This, together with the OPE {\aii}, forces the field ${\Sigma}$ to
be a free fermion type theory \ref\fj{P.~Federbush and K.~Johnson,
Phys.\ Rev {\bf 120}, 1926 (1960).}

This means, the form of the OPE of the field ${\Sigma}$, can be
written down as follows:
\eqn\aii{{\Sigma(z)} {\Sigma(0)}  {\sim}   z^{-1}  {1} + z  {\cal
O}(0) + ... }

In {\aii},
 ${\cal O}$ is the stress tensor of the free fermion
theory, and is an operator of dimension 2, which we will associate
with the dimension 2 operator that appears in the extended
superconformal algebra of {\sv}.The OPE of ${\Sigma}$ with the
superconformal generator T$_F$, is determined by the BRST invariance
of the gravitino vertex, and forces the OPE to have terms no more
singular than $z^{-1/2}$.That is,

\eqn\aiii{ {\Sigma(z)} T_F (0)  \sim  z^{-1/2}}

Using the above OPE and the identification of the spin field
${\Sigma}$ with the theory of free majorana fermions, one arrives at
the following OPE between ${\cal O}$ and T$_F$ :

\eqn\aiv{ {{\cal O}(z)} T_F (0)  \sim  z^{-2} A(0)  +  z^{-1} {\cal
W }(0) + ... }

Dimensional analysis tells us that $A$ has dimension 3/2, while
${\cal W}$ has dimension 5/2. This is the same as the dimension 5/2
operator that appears in {\sv}. As for $A$, since it has dimension
3/2 we will identify it with the supercurrent T$_F$.

This identification can easily be checked by referring back to the
free fermion picture, where an explicit expression can be written
down for ${\cal O}$ in terms of ${\Sigma}$ , using which one finds
the OPE {\aiv} with the identification of A with T$_F$ as we
claimed.More generally, as noted in \ref\zam{A.~B.~Zamolodchikov,
``Infinite Additional Symmetries In Two-Dimensional Conformal
Quantum Field Theory,'' Theor.\ Math.\ Phys.\  {\bf 65}, 1205 (1985)
[Teor.\ Mat.\ Fiz.\ {\bf 65}, 347 (1985)]} , a symmetry current of
weight 3/2 can only correspond to the supercurrent, in a unitary
conformal field theory \footnote{$^2$} {A way to see this is to note
that, if the spin ${3\over 2}$ current is not the super-current, its
OPE with the super-current forces the introduction of a
super-partner, either of spin(1) or of spin(${5 \over 2}$). The
spin(1) current together with the rest of the symmetry currents
forms the ${\cal N}=2$ SCA(Calabi-Yau situation), whereas, with
spin($2$), the only way to close the OPE is with $c={21 \over
2}$(this is the G$_2$ case), or we can append the G$_2$ algebra with
a free $c=1$ theory, which is nothing but a Spin(7) CFT again, as
shown in \ref\gepn{D.~Gepner and B.~Noyvert, ``Unitary
representations of SW(3/2,2) superconformal algebra,'' Nucl.\ Phys.\
B {\bf 610}, 545 (2001) [arXiv:hep-th/0101116].}(geometrically this
can be thought of as Spin(7)s which are S$^1$ fibered over a G$_2$
holonomy manifold).}. So, the OPE {\aiv} can be re-written as
follows:

\eqn\avi{ {{\cal O}(z)} T_F (0)  \sim  {3 \over 4c} {T_F(0)\over
{z^2} } + z^{-1} {\cal W }(0) + ... }
In order to confirm the existence of this algebra, the OPEs between
T, T$_F$ ,${\cal O}$ and ${\cal W}$ , must close without
introduction of further operators. The algebra discussed above
closes only if $c=12$, and the rest of the OPEs can be identified by
the requirement of closure of the algebra \footnote{$^{\dag}$} {This
algebra has been studied previously, in {\gepn} and was first
derived in \ref\fig{J.~M.~Figueroa-O'Farrill and S.~Schrans,
``Extended Superconformal Algebras,'' Phys.\ Lett.\ B {\bf 257}, 69
(1991). }}. Indeed the ${\cal SW}({3\over 2},2)$ algebra is the
unique ${\cal N}=1$ SCA containing precisely one spin-2 multiplet in
addition to the stress tensor and supercurrent of a generic ${\cal
N}=1$ SCA. Furthermore for $c=12$ the ${\cal SW}({3\over 2},2)$
algebra contains an Ising subsector {\gepn} which is what forced
$c=12$ in our case.

In the above equations, $c$ refers to the central charge of the
internal ${\cal N} = 1 $ super-algebra, which is determined from the
requirement of the theory providing a good string background, to be
$c = 12$, which is the same requirement for the closure of this
algebra.To be more precise, the operator ${\Sigma}$ was our
spin-field, which created the Ramond ground state in the internal
${\cal N} =1$ SCFT on ${\cal N}$, and this being the case,
${\Delta}[{\Sigma}] = {c\over 24} $ and hence the central charge of
the internal SCFT is required to be 12. As noted, this is also the
same value of the central charge, for which the extended algebra
discussed above closes. Most of the proof in the above section
assumed that the extended super-conformal algebra that arises from
the requirement of ${\cal N} = 1 $ supersymmetry in the target
space, was minimal. It is in fact a generic situation for spin(7)
manifolds, though we will argue later that for certain singular
spin(7) manifolds there are additional conserved currents on the
worldsheet.

Suppose we have ${\cal N} =2$ supersymmetry in the external
space-time, one should recover the extended super-conformal algebra
underlying Calabi-Yau compactifications. It is easy to see that this
is indeed the case. With twice as much supersymmetry in the target
space, we have two spin-fields ${\Sigma}_1$ and ${\Sigma}_2$, which
satisfy the equations:

\eqn\ax{  {\Sigma}_I(z)  {\Sigma}_J(0) \sim  {\delta}_{IJ}{1 \over
z} + ...}

This means, we have two free majorana fermions, so we can use the
usual bosonization prescription, and conclude that the theory
possesses a spin(1) conserved current, call it $J$. Then, the
existence of this spin(1) current means, as usual that the ${\cal
N}=1$ SCA on the world-sheet is enhanced to a ${\cal N} =2$ SCA. The
spin-field can be written in terms of the bosonized R-current as
${\Sigma}_1 + i{\Sigma}_2  = e^{i{\phi}}$, where $J =
i{\sqrt{4}{\partial}{\phi}}$.

Actually, the extended superconformal algebra associated with
Calabi-Yau manifolds is in fact larger and contains as a subset the
${\cal N}=2$ SCA. This larger algebra is generated by including the
spectral flow generator and its ${\cal N}=2$ superpartners. It can
be easily checked the a ${\bf Z}_2$ projection of this larger
algebra leaves a closed algebra of the form ${\cal SW}({3\over
2},2)$ and is simply the statement that there are Spin(7) spaces
that are obtained by orbifolding Calabi-Yau four-folds.

\newsec{Singularities}

It is familiar in the context of Calabi-Yau spaces, that the
worldsheet theory becomes singular as certain K{\"a}hler and
complex-structure parameters are tuned. At these points in moduli
space, the worldsheet CFT acquires a ``throat'' and the physics down
the throat is strongly coupled and describable in many cases as a
Liouville type SCFT. The worldsheet CFT being strongly coupled
coincides with certain non-perturbative objects like D-branes
wrapping supersymmetric cycles becoming light. A very concrete
prescription for the non-critical string theory describing a class
of Calabi-Yau singularities was given in \ref\GiveonZM{
  A.~Giveon, D.~Kutasov and O.~Pelc,
  ``Holography for non-critical superstrings,''
  JHEP {\bf 9910}, 035 (1999)
  [arXiv:hep-th/9907178]}. The worldsheet description of
  non-critical ${\cal N}=1$ strings requires a linear dilaton
  multiplet tensored with a suitable  SCFT ${\cal N}$. The linear
  dilaton multiplet consists of a scalar with background charge
  ${\phi}$ together with a Majorana fermion ${\psi}_{{\phi}}$.
  In order to describe a string propagating on a spin(7) manifold we
  expect the internal CFT (the linear dilaton together with ${\cal
  N}$ should provide a ${\cal SW}({3\over 2},2)$ SCA. We shall argue
  that de-coupling the linear dilaton multiplet from this enhanced
  SCA gives rise to a ${\cal SW}({3\over 2},{3\over 2},2)$ algebra.
  In other words, a non-critical string that preserves two
  supercharges, can be constructed by tensoring the ${\cal N}=1$
  linear dilaton CFT with a matter CFT with the enhanced SCA ${\cal
  SW}({3\over 2},{3\over 2},2)$, with $c = {{21}\over 2} - 3Q^2$,
  where $Q$ is the background charge of the linear dilaton.

The ${\cal N}=1$  ${\cal SW}({3\over 2},{3\over 2},2)$ algebra is
generated by the stress tensor $T$, the supercurrent $G$, a
spin-${3\over 2}$ super multiplet and a spin-2 super multiplet. This
algebra exists for all $c$ and a parameter ${\l}$ called the
self-coupling. For $c={{21}\over 2}$ this is the $G_2$ algebra
studied in {\sv}. For more details about the representation theory
and the commutation relations that define the algebra we refer the
reader to {\Noyvert}.

  The
  straightforward way to show the claim is simply to start with a
  free-field representation of a ${\cal SW}({3\over 2},2)$ algebra
  extended by a dimension $1\over 2$ multiplet
  $({\psi}_{\phi},{\phi})$ where ${\phi}$ carries background charge
  and show that one can de-couple this multiplet resulting in a
  ${\cal SW}({3\over 2},{3\over 2},2)$ SCA at $c = {{21}\over 2}
  -3Q^2$.

In fact a calculation very similar to this has already been done in
{\gepn} where the authors showed that a ${\cal SW}({3\over 2},2)$
algebra at $c=12$ containing a $h={1\over 2}$ multiplet could be
re-written after de-coupling the free fermion and free current as
the $c={3\over 2}$ free theory tensored with a ${\cal SW}({3\over
2},{3\over 2},2)$ SCA at $c= {{21}\over 2}$. The only difference in
our case is that the $h={1\over 2}$ multiplet is not free. With
minor changes their proof still goes through, showing that the
tensor product of a ${\cal N}=1$ linear dilaton together with ${\cal
SW}({3\over 2},{3\over 2},2)$ algebra generates a spin(7) algebra
which is extended.

   We will however provide alternate
arguments which we believe are more intuitive.
   We start with the observation
  that a large class of singular Calabi-Yau manifolds can be
  orbifolded to give rise to singular spin(7) geometries. We know
  that the non-critical string describing the singular Calabi-Yau geometries is a
  (deformation of) the ${\cal N}=2$ linear dilaton CFT tensored with
  an internal ${\cal N}=2$ SCFT. The orbifolding involves $J
  \rightarrow -J$. Our claim then amounts to the statement that the
  ${\bf Z}_2$ orbifold of the tensor product of $S^1$ with an ${\cal
  N}=2$ SCA gives rise to a ${\cal SW}({3\over 2},{3\over 2},2)$
  SCA.

One can view this as the CFT generalization of the familiar
statement that a ${\bf Z}_2$ orbifold of a circle fibration of a
Calabi-Yau three-fold gives rise to a $G_2$ holonomy manifold.

Since our analysis is going to be similar to the more familiar $G_2$
case, we will recall how the argument goes in the simpler setting.
We know that a class of $G_2$ holonomy spaces can be obtained by
considering ${{{S^1 \times CY_3}}\over {{\bf Z}_2}}$. The worldsheet
description of $CY_3$ compactifications is in terms of a ${\cal
N}=2$ SCFT with $c =9$ and integral R-charges. As usual, let us
bosonize the U(1) R-current by $J = i{\sqrt{3\over
2}}{\partial}{\phi}$. The two ${\cal N}=2$ super currents are
denoted by $G^{\pm}$, and the stress tensor by $T$. The $S^1$ part
is simple enough, and given by a free fermion ${\psi}_{\th}$
together with ${\partial}{\th}$. The ${\bf Z}_2$ orbifold acts by
${\psi}_{\th} \rightarrow -{\psi}_{\th}$, ${\th} \rightarrow -{\th}$
and ${\phi}\rightarrow -{\phi}$. Tensoring the two CFTs together, is
it possible to identify a ${\cal SW}({3\over 2},{3\over 2},2)$
algebra that is left invariant under the projection? The candidate
for the spin-${3\over 2}$ generator is:
\eqn\ci{H = e^{i{\sqrt 3}{\phi}} + e^{-i{\sqrt 3}{\phi}} +
:J{\psi}_{\th}: }

Taking OPEs with the super-current $G = G^+ + G^- +
{\psi}_{\th}{\partial}{\th}$, yields the spin-2 partner. The other
generators are obtained from the $HH$ and $HM$ OPEs. The full set of
generators form the $G_2$ superconformal algebra as they should.

Analogously singular CY 4-folds are obtained by tensoring the ${\cal
N}=2$ linear dilaton CFT (at $c = 3 + 3Q^2$) together with a ${\cal
N}=2$ SCFT at $c= 9-3Q^2$. GSO projection allows the presence of the
following operator:
\eqn\cii{ U =: e^{iQ{\th}} e^{i{\sqrt{c\over 3}}{\phi}}: +
e^{-iQ{\th}} e^{-i{\sqrt{c\over 3}}{\phi}}: }

which has spin ${3\over 2}$ and survives GSO projection. So the most
general spin ${3\over 2}$ operator is of the form $H = U +
:J{\psi}_{\th}: + G$. Inclusion of this operator is expected to lead
to the ${\cal SW}({3\over 2},{3\over 2},2)$ algebra at $c={{21}\over
2}-3Q^2$ upon orbifolding the $U(1)$ R-current by $J \rightarrow -J$
to go from CY 4-folds to spin(7)s. This orbifolding is the
worldsheet description of the anti-holomorphic involution that gives
us a spin(7) starting from a Calabi-Yau four-fold.

The ${\cal SW}({3\over 2},{3\over 2},2)$ algebra is determined in
terms of two parameters the central charge $c$ and the coupling
${\l}$. In fact the algebras of relevance to singular spin(7)
manifolds must have ${\l}$ fixed. Indeed it was shown by Noyvert
that there is a real ${\cal N}=1$ SCA with $c = {7\over {10}}$ for
any value of central charge $c \geq {3\over 2}$ if $\l$ obeys:
\eqn\ciii{ {\l} = {4(63-6c)(6c-9)^2}\over {243(30c - 21)} }
 Of course, this real ${\cal N}=1$ SCA is that of the tri-critical
 Ising model. The existence of this tri-critical Ising algebra
 allows us to decompose the conformal blocks of the internal ${\cal
 N}=1$ SCFT such that we can preserve supersymmetry\footnote{$^2$}{ Indeed the fact
that the presence of a Tri-critical Ising sector allows one to write
a supersymmetric partition function was independently noticed by
Eguchi et.al
  \ref\Eguchi{
  T.~Eguchi, Y.~Sugawara and S.~Yamaguchi,
  ``Supercoset CFT's for string theories on non-compact special holonomy
  manifolds,''
  Nucl.\ Phys.\ B {\bf 657}, 3 (2003)
  [arXiv:hep-th/0301164].}. Our paper can be thought of
  as providing the conditions under which a Tri-critical Ising algebra exists in a spin(7) compactification.}.

 Now ${\l}$ vanishes for $c = {{21}\over 2}$ where we recover the
 $G_2$ SCA. It also vanishes for $c = {3\over 2}$. This puts $Q =
 {\sqrt{3}}$. The ${\cal N}=1$ SCFT with this enhanced SCA is
 nothing but the ${\bf Z}_2$ orbifold of the free boson on a circle
 of radius ${\sqrt 3}$ together with a free boson. The orbifold acts
 by ${\th} \rightarrow -{\th}$ and ${\psi}_{\th} \rightarrow -
 {\psi}_{\th}$. Precisely this model arises upon decoupling the
 Liouville field from ${\cal N}=2$ super-Liouville theory with
 $c=12$. In fact it is GSO projection that fixes $Q = {\sqrt 3}$ in
 this case.

 Let us consider the ${\cal N}=2$ super-Liouville theory with
 $c=12$. Asymptotically the theory is formulated in terms of a
 free ${\cal N}=1$ SCFT with a compact boson ${\th}$ and free Majorana
 fermion ${\psi}_{\th}$ together with the ${\cal N}=1$ linear dilaton
 system comprising ${\phi}$ (the Liouville field) and its
 superpartner ${\psi}$. This background is a singular Calabi-Yau
 4-fold CFT if $c=12$. Indeed the corresponding non-compact CY is
 nothing but the affine variety:
 \eqn\di{ z_1^2 + z_2^2 + z_3^2 + z_4^2 + z_5^2 = 0 \in {\bf C}^5 }

 The deformation of this variety to make it non-singular corresponds
 to turning on a constant in the RHS of {\di}. It is well known
 that the effect of turning on ${\mu}$ is the same as turning on the
 worldsheet cosmological constant of the ${\cal N}=2$
 super-Liouville theory to which the non-compact CY {\di} is mirror
 dual. The worldsheet cosmological constant is:
 \eqn\dii{ e^{-{{\varphi}}} e^{{i\over Q}{\th}}e^{{1\over Q}{\phi}} }

 written in -1 picture. For $c=12$, $Q = {\sqrt 3}$ which implies
 ${\th}$ lives on a circle of radius ${\sqrt 3}$. Upon quotienting
 by the ${\bf Z}_2$ anti-holomorphic involution ${\th} \rightarrow
 -{\th}$, ${\psi}_{\th} \rightarrow -{\psi}_{\th}$ the free
 $c={3\over 2}$ CFT of ${\th}$ and ${\psi}_{\th}$ becomes precisely
 the ${\cal N}=1$  ${\cal SW}({3\over 2},{3\over 2},2)$ SCFT at
 $c={3\over 2}$.

 One of the examples treated in {\Eguchi} was a coset based on
 $SO(7)/G_2$ where it was argued to correspond to a noncompact
 spin(7) SCFT. Using the structure constants $f_{abc}$ of the
 octonions, it is easy to see that there exist holomorphic
 spin-${3\over 2}$ current $U = f_{abc}{\psi}^a {\psi}^b{\psi}^c$
 and a spin-2 current $V =
 *f_{abcd}{\psi}^a{\psi}^b{\psi}^c{\psi}^d$ which are conserved,
 where ${\psi}^a$ are the seven free fermionic superpartners of the
 coset. Together with the ${\cal N}=1$ generators the currents
 $(U,V)$ generate the ${\cal SW}({3\over 2},{3\over 2},2)$ algebra.

The ${\cal N}=2$ supersymmetric worldsheet cosmological constant is
invariant under the ${\bf Z}_2$ projection and gives us a ${\cal
N}=1$ superpotential which is invariant under the spin(7) algebra.
Interestingly enough, this is not the only deformation that one can
do. In fact there exists a ${\cal N}=1$ supersymmetric deformation
that also preserves the spin(7) SCA but does not arise by ${\bf
Z}_2$ projection of a ${\cal N}=2$ potential. This corresponds in
the geometric terms to deforming away from the limit where the
spin(7) has ${\pi}_1 ={\bf Z}_2$ into a noncompact spin(7) with
trivial fundamental group. The relationship between this SCFT and
the corresponding noncompact spin(7) manifold will be discussed
elsewhere.

Even though we have motivated the existence of the $G_2$ like
enhanced SCA, it would be nice to get a simpler explanation as to
why there is such an extended algebra for singular spin(7)s. In the
next section we show that vacua of linear dilaton (or $AdS_3$) type
that have two supercharges also end up giving rise to conserved
worldsheet currents in the standard way. The basic observation here
is that starting with a singular spin(7) of the form ${\bf R}_{\phi}
\times {\cal N}$ and adding F-strings takes us to a background of
the form $AdS_3 \times {\cal N}$. One can then use supergravity to
constrain the type of ${\cal N}$ that gives rise to supersymmetric
vacua. This constraint on the geometry of ${\cal N}$ can be re
phrased as the existence of certain holomorphic currents in the CFT
describing string propagation on ${\cal N}$.

\newsec{ $AdS_3$ vacua that preserve two supercharges}

Starting from the linear dilaton vacua discussed previously, one can
construct vacua of $AdS_3$ type, by adding fundamental strings to
vacua of the form ${\bf R^{1,1}} \times {\bf R}_{\phi} \times {\cal
N}$. Considering a large number $p$ of F1-strings spanning ${\bf
R}^{1,1}$, we end up with $AdS_3 \times {\cal N}$. On the
worldsheet, $AdS_3$ is described by a ${\bf SL(2,R)}$ Kac-Moody
algebra at level $k$, so that taking $k$ large the sigma model
describing both $AdS_3$ and ${\cal N}$ becomes weakly coupled and we
expect supergravity to be a good approximation. In the supergravity
limit, we shall examine the conditions for solutions of the form
$AdS_3 \times {\cal N}$ to be supersymmetric vacua, and determine
constraints on ${\cal N}$ that arise this way. As we show below, the
supergravity analysis directly reveals the presence of a holomorphic
spin-${3\over 2}$ and spin-2 operator, which together with their
${\cal N}=1$ super partners give us the field content of a ${\cal
SW}({3\over 2},{3\over 2},2)$ algebra.

Our starting point is an ansatz for the supergravity background of
the form:
\eqn\ei{ds^2 = (AdS_3) + g_{mn}dx^mdx^n }
In {\ei} $g_{mn}$ refers to the metric on ${\cal N}$.

In order for part supersymmetry to be preserved in this background,
the gravitino and dilatino variations must vanish. We are looking
for backgrounds that are created by NS5-branes and F-strings only,
so the only fields that are turned on are the metric and 3-form
NS-NS flux $H$. In order for the gravitino variation to vanish, we
must have \footnote{$^4$} {The spinors ${\e}^{\pm}$ are 10d
Majorana-Weyl spinors of the same (opposite) chirality for type
IIB(A)}:
\eqn\eii{ {{\nabla}_M}^{\pm} {\e}^{\pm} = {\nabla}_M - {1\over
8}H_{MNP}{\G}^{NP}{\e}^{\pm} = 0 }

For the dilatino variation to vanish, we need:
\eqn\eiii{ H^{MNP} {\G}_{MNP}{\e}^{\pm} = 0}

In {\eii} and {\eiii} the indices $M,N,P$ refer to the 10
dimensional space-time while lower case indices $m,n,p$ refer to
coordinates on ${\cal N}$ and greek indices ${\mu},{\nu}$
parameterize $AdS_3$. The dilaton is assumed to be constant for our
background so we have not shown its dependence explicitly.

Respecting the 3 ${\oplus}$ 7 split of the metric, the ten
dimensional spinors ${\e}^{\pm}$ can be split as:
\eqn\evi{ {\e}^{\pm} = {\eta} \otimes {\xi}^{\pm} }

The spinor ${\eta}$ is a conformal Killing spinor of $AdS_3$, so it
satisfies the following equation:
\eqn\evii{ {\nabla}_{\mu} {\eta} = {k\over 2}{\g}_{\mu}{\eta} }

The constant $k$ in {\evii} is related to the cosmological constant
of $AdS_3$. Since ${\cal N}$ is a seven dimensional Euclidean
manifold, the spinors ${\xi}^{\pm}$ are Majorana. This means that in
order to have two supercharges in $AdS_3$ (actually four including
the superconformal generators), there must be precisely two nowhere
vanishing spinor ${\xi}^{\pm}$ in the internal space ${\cal N}$.

The ansatz for the H-flux consistent with maximal symmetry of
$AdS_3$ is:
\eqn\eviii{H_{{\mu}{\nu}{\rho}} = f{\e}_{{\mu}{\nu}{\rho}}  }

The mixed components of $H$ have to vanish, while $H_{mnp}$ is
arbitrary, subject to conditions following from {\eii} and {\eiii}.
In {\eviii}, $f$ measures the NS5-brane charge of the background.

In order for {\eiii} to hold, we require:
\eqn\eix{ {H \!\!\!\!\!\, /} {\xi}^{\pm} = {\pm}6f{\xi}^{\pm} }

where we used the contraction ${H \!\!\!\!\!\, /} =
H_{mnp}{\g}^{mnp}$. This, along with {\eii} tells us that the spinor
${\xi}$ must be covariantly constant with respect to a connection
with torsion:
\eqn\ex{ {\nabla}_m{\xi}^{\pm} {\mp} 12e^{-{1\over 2}{\phi}}
H_{mnp}{\g}^{np}{\xi}^{\pm} = 0}

which can be written as:
\eqn\exa{ {\nabla}_{\pm} {\xi}^{\pm} = 0 }

The existence of a spinor ${\xi}^+$ satisfying {\exa} guarantees the
existence of the three form ${\Phi}_{mnp} = {{\xi}^+}^T {\g}_{mnp}
{\xi}^+$, and the four-form $*{\Phi}$ that are also
$H^+$-covariantly constant.

The worldsheet description of this string background is a ${\cal
N}=(1,1)$ supersymmetric non-linear sigma model, whose action is:
\eqn\exi{S= \int {d^2z (G_{ij}({\phi}) +
B_{ij}({\phi})){\partial}{\phi}^i{\bar {\partial}}{\phi}^j +
{\psi}_i({\bar {\partial}}{\psi}^i + {{{\G}_+}^i}_{jk}{\bar
{\partial}}{\phi}^j{\psi}^k)}}

where $B$ is the NS-NS 2-form field whose field strength is $H$.
${\phi}^i$ are coordinates on ${\cal N}$ and ${\psi}^i$ are the
fermionic superpartners. As the background is a classical solution
of string theory, the full action is order by order conformally
invariant. However, the fields ${\psi}^i$ and ${\phi}^i$ no longer
obey the free-field equations. All OPEs acquire corrections that
depend on the curvature of ${\cal N}$ and depend on the parameter
$f$. For $f\rightarrow 0$ we must recover the OPEs of the $G_2$
conformal algebra as in {\sv}.

By using the classical equations of motion for the ${\psi}$ field,he
worldsheet operators $U = {\Phi}_{mnp}{\psi}^m{\psi}^n{\psi}^p$ and
$X =
*{\Phi}_{mnpq}{\psi}^m{\psi}^n{\psi}^p{\psi}^q$ are seen to be holomorphic.
From their definition, it is apparent that $U$ has spin-${3\over 2}$
and $V$ has spin-2\footnote{$^5$}{This is true even though
${\psi}^i$ do not have canonical OPEs, as the spin of the fields are
independent of $k$.}. Together with their ${\cal N}=1$
superpartners, they have the right field content to form a ${\cal
SW}({3\over 2},{3\over 2},2)$ algebra. Similarly there is an
anti-holomorphic sector of conserved higher spin currents also (the
theory is parity invariant upon $B \rightarrow -B$).

It has in fact been noticed by Giveon and Rocek \ref\GiveonJG{
  A.~Giveon and M.~Rocek,
  ``Supersymmetric string vacua on AdS(3) x N,''
  JHEP {\bf 9904}, 019 (1999)
  [arXiv:hep-th/9904024].}
from the worldsheet point of view, that orbifolding by the $U(1)$
R-current of the internal ${\cal N}=2$ SCA of a string vacuum of the
form $AdS_3 \times S^1 \times {\cal M}$ gives rise to an $AdS_3
\times {\cal N}$ vacuum with ${\cal N}=1$ supersymmetry. In that
paper, it was also speculated that the internal SCFT corresponding
to ${\cal N}$ might have an enhanced SCA. The result of this paper
are in complete agreement with the expectations of {\GiveonJG}.

Furthermore, if we now have one more pair of Majorana spinors
solving {\eix} and {\exa} it is easy to show that they define a
nowhere vanishing vector field such that one can write ${\cal N} =
S^1 \times {\cal M}$ with ${\cal M}$ a K{\"a}hler manifold with
torsion such that there are two-forms $J = {{\xi}^+}^T_1 {\g}^{ab}
{{\xi}^+}_2$ and ${\bar J} = {{\xi}^-}^T_1 {\g}^{ab} {{\xi}^-}_2$
which are covariantly constant with respect to $H^+$ and $H^-$
connections. This automatically leads to the worldsheet theory being
described by a $U(1)$ fibration of a ${\cal N}=2$ SCFT, which was
found independently from the worldsheet perspective in {\GiveonJG}.

\newsec{Discussion}

In this note we have pointed out that ${\cal N}=1$ linear dilaton
CFT can be tensored with a ${\cal SW}({3\over 2},{3\over 2},2)$ SCFT
and GSO projected to form a consistent background which preserves
two space-time supercharges. It is natural to expect that these
vacua are related to singular limits of spin(7) CFTs, and indeed
their construction shows the appearance of the enhanced SCA of
spin(7)s naturally. It would be very interesting to obtain more
examples of ${\cal SW}({3\over 2},{3\over 2},2)$ SCFTs in light of
this fact. Moreover, there is a canonical way to resolve these
singularities in the worldsheet theory by turning on a potential
that is invariant under the spin(7) algebra. This would describe
string propagating on certain noncompact spin(7) manifolds. It is
quite non-trivial to identify the precise relationship between a
given CFT and the geometric model it describes. In the ${\cal N}=2$
situation remarkable simplicity arises because of the well known
connection\ref\MartinecZU{
  E.~J.~Martinec,
  ``ALGEBRAIC GEOMETRY AND EFFECTIVE LAGRANGIANS,''
  Phys.\ Lett.\ B {\bf 217}, 431 (1989).}\ref\VafaUU{
  C.~Vafa and N.~P.~Warner,
  ``CATASTROPHES AND THE CLASSIFICATION OF CONFORMAL THEORIES,''
  Phys.\ Lett.\ B {\bf 218}, 51 (1989).} between Landau-Ginzburg models and Calabi-Yau spaces realized as
algebraic varieties in projective/ weighted projective spaces (which
are thought of as defining the superpotential data of a ${\cal N}=2$
supersymmetric UV free theory which flows to the corresponding
Calabi-Yau sigma model in the IR). An analogous simplification does
not seem to exist for spin(7) spaces. It would be very interesting
to determine the analog of LG/ CY correspondence for spin(7) spaces.
A typical ${\cal N}=1$ supersymmetric LG theory does not have
protected operators, in particular the superpotential does not
parameterize the IR SCFT to which it conjecturally flows. Perhaps in
the spin(7) case the existence of the enhanced SCA might persist in
some form even in the UV, which allows one to make
non-renormalization type statements.

By adding F1-strings we were able to show that the $AdS_3$ vacua
which preserve ${\cal N}=1$ supersymmetry all have an enhanced SCA
on the worldsheet of the form ${\cal SW}({3\over 2},{3\over 2},2)$.
This completes the classification of $AdS_3$ vacua with
NS-flux\ref\GiveonKU{
  A.~Giveon and A.~Pakman,
  ``More on superstrings in AdS(3) x N,''
  JHEP {\bf 0303}, 056 (2003)
  [arXiv:hep-th/0302217].}.

An entirely analogous discussion can be carried out to construct
singular $G_2$ SCFTs. These SCFTs would not suffer from the 1-loop
destabilization \ref\VafaFJ{
  C.~Vafa and E.~Witten,
  ``A One Loop Test Of String Duality,''
  Nucl.\ Phys.\ B {\bf 447}, 261 (1995)
  [arXiv:hep-th/9505053].}
 that affects compactifications on eight dimensional
manifolds and forces turning on RR-flux when the tadpole for
$B$-field does not vanish in type IIA. Indeed the tadpole condition
vanishes identically for seven dimensional manifolds allowing us to
consider compactifications without RR fluxes.

\newsec{Acknowledgements}

I would like to acknowledge Markus Luty and my advisor James
Sylvester Gates  for constant encouragement and support, and for
numerous enlightening discussions.

 \listrefs

\end